\documentclass[a4paper]{article}
\usepackage{a4wide}
\usepackage[normalem]{ulem}
\usepackage{dsfont}
\usepackage[T1]{fontenc}
\usepackage{amsmath}
\usepackage{amssymb}
\usepackage{amsthm}
\usepackage[ansinew]{inputenc}
\usepackage{enumerate}
\usepackage{fancyhdr}
\usepackage{color}
\usepackage{ifpdf}
\usepackage{rotating}
\usepackage[T2A]{fontenc}
\usepackage{mathtext}

\DeclareMathOperator{\diag}{diag}
\newtheorem{deff}{Definition} 
\newtheorem{thm}{Theorem}
\newtheorem{prop}{Proposition}
\newtheorem{cor}{Corollary}

\begin{document}
 \title{Future non-linear stability for reflection symmetric solutions of the Einstein-Vlasov system of Bianchi types II and VI$_0$}
\author{Ernesto Nungesser}   
\maketitle
\begin{abstract}
 Using the methods developed for the Bianchi I case we have shown that a boostrap argument is also suitable to treat the future non-linear stability for
reflection symmetric solutions of the Einstein-Vlasov system of Bianchi types II and VI$_0$. These solutions are asymptotic to the
Collins-Stewart solution with dust and the Ellis-MacCallum solution respectively. We have thus generalized the results obtained by Rendall 
and Uggla in the case of locally rotationally symmetric Bianchi II spacetimes to the reflection symmetric case. 
However we needed to assume small data. For Bianchi VI$_0$ there is no analogous previous result. 
\end{abstract}
\pagenumbering{roman}
\tableofcontents
\newpage

\pagenumbering{arabic}

 \section{Introduction}
  A starting point to understand general cosmological models are the homogeneous models. There are a lot of results concerning this subject and 
in particular two books \cite{RS}, \cite{WE} which are an excellent introduction and a great summary
 of many of the results obtained. 

In general the focus has been on the fluid model since it appears (theoretically)
 relatively natural when dealing with isotropic universes and from observations we also know that 
the Universe is almost isotropic. However to have a deeper understanding of the dynamics one should
 go beyond the study of isotropic universes. General statements may vary then depending on the 
choice of the matter model. It is also important to note as is pointed out in \cite{EW} that 
a quasi-isotropic epoch is compatible with all Bianchi models and thus it is interesting
 to study the dynamics of all the different types.

We will deal with the \emph{future asymptotics} of some homogeneous cosmological models within
 the so called Bianchi class A and the matter is described via an ensemble of free falling particles 
also called collisionless matter. For all the models treated here the fundamental questions are on a 
firm ground, i.e. future geodesic completeness has been shown for these models \cite{CC}, \cite{GC}.

Concerning the late time behaviour of the Universe one believes in the cosmic no hair conjecture. This conjecture states roughly 
speaking that all expanding cosmological models with a positive cosmological constant approach asymptotically the de Sitter solution.
 For the Einstein-Vlasov system isotropization could be shown for all non-type IX Bianchi
 cosmologies \cite{Lee}. 

In absence of a cosmological constant there are also different results concerning the future.
For the Einstein-non-linear scalar field system we refer to \cite{HR1}, \cite{HR2}, \cite{Alho} and \cite{Costa}.
The late-time behaviour of Bianchi spacetimes with a non-tilted fluid is well understood \cite{JW}, \cite{HW}. 
In particular all non-tilted perfect fluid orthogonal Bianchi models except IX with a linear equation of
 state where $0<\gamma<\frac{2}{3}$ are future asymptotic to the flat FL model \cite{HUW}. For other values of $\gamma$ one cannot expect
 isotropization for most of the Bianchi models. However there is an important characteristic of the future asymptotics for the 
Bianchi types I, II and VI$_0$, namely that the spacetimes considered tend to special (self-similar) solutions. For expanding models it is reasonable
to expect that the dispersion of the velocities of the particles will decay. The conjecture now is that for the Einstein-Vlasov system the spacetimes 
of certain Bianchi type will tend to the same special solution as for the corresponding Einstein-dust system.

This has been achieved already for locally
 rotationally symmetric (LRS) models in the cases of Bianchi I, II, III, VIII and IX \cite{RT},\cite{RU},\cite{RE}, \cite{Gernot}, \cite{SI}
 and for reflection symmetric models in the case of Bianchi I \cite{IS}. These results have been obtained 
using dynamical systems theory.

The Einstein-Vlasov system remains a system of partial differential equations (PDE's)
 even if one assumes spatial homogeneity. The reason is that although the distribution function
 written in a suitable frame will not depend on the spatial point, the dependence with respect
 to the momenta remains (since the Vlasov equation is defined on the mass shell). However in
 the results mentioned a reduction to a system of ordinary differential equations was possible due
 to the additional symmetry assumptions. This is no longer possible if one drops some of these
 additional symmetries (see \cite{MM2} for the reasons). Thus if one wants to generalize 
these results the theory of finite dimensional dynamical systems is not enough.

Most of the results obtained until now rely on the theory of dynamical systems.
Thus one might be tempted to use techniques coming from the theory of infinite-dimensional dynamical systems.
 The first important difficulty would be to choose the suitable (weighted) norm. Another one is that 
important theorems which have been used for the finite-dimensional case cannot be used here. All this may
 work, but this is not the approach taken here.

Here the main tool used is a bootstrap argument which is often used in non-linear PDE's.
 We will present results concerning the late-time behaviour of some expanding Bianchi A spacetimes with
 collisionless matter where we have assumed small data. This assumption will be specified later, but 
roughly it means that the universe is close to the special self-similar solution mentioned earlier and that
 the velocity dispersion of the particles is small.

The results obtained are as follows. For reflection symmetric Bianchi II
 and reflection symmetric Bianchi VI$_0$ we have been able to show that their late-time behaviour remains
 the same if the LRS condition is dropped. We will show that these spacetimes, reflection symmetric
 Bianchi II and reflection symmetric Bianchi VI$_0$, will tend to solutions which are even more symmetric.
In the case of Bianchi II we will show that it will become LRS, a Bianchi model whose isometry group of
 the spatial metric is four-dimensional. In this case there exists a one-dimensional isotropy group
 and one can show that a spacetime of Bianchi class A admits a four-dimensional isometry group, if and
 only if two structure constants are equal and if the corresponding metric components are equal as well. 
Bianchi VI$_0$ cannot be LRS, however it is compatible with an additional discrete symmetry
 (Appendix B.1 of \cite{CH}). The analysis of the asymptotics shows that the Bianchi VI$_0$ spacetimes
 tend to this special class. Note that for VI$_0$ there is no corresponding LRS/previous result. 

To be more precise we show that the reflection symmetric solutions of Bianchi type II and VI$_0$ are asymptotic to the
Collins-Stewart solution with dust and the Ellis-MacCallum solution respectively. The asymptotic behaviour of the metric and the
energy-momentum tensor is studied in detail.

All the results show that the dust model usually assumed in observational cosmology in the
 'matter-dominated' Era is robust. Another way of saying the same is that asymptotically collisionless
 matter is well approximated by the dust system.

The paper is organized as follows. In the following section we will present the general basic
 equations, namely the Einstein-Vlasov system. In sections three and four we explain the symmetry assumptions
 and deduce the corresponding equations of the Einstein-Vlasov system which are then summarized in section
 five. Afterwards we present some special solutions and their linear stability as a pre-stage of our main
 argument in section eight: the bootstrap argument. This argument is refined in section nine and leads to
 our main results. In our last section we have a discussion about our results and
 present some possible future directions.

\section{Relativistic kinetic theory}
\subsection{The Einstein-Vlasov system}
A cosmological model represents a universe at a certain averaging scale. It is described via a Lorentzian metric $g_{\alpha\beta}$
 (we will use signature -- + + +) on a manifold $M$ and a family of fundamental observers. The metric is assumed to be time-orientable, 
which means that at each point of $M$ the two halves of the light cone can be labelled past and future in a way which varies continuously
 from point to point. This enables to distinguish between future-pointing and past-pointing timelike vectors. This is a physically
 reasonable assumption from both a macroscopic point of view e.g. the increase of entropy and also from a microscopic point of view e.g.
 the kaon decay.
As we already mentioned in the introduction one has also to specify the matter model and this we will do in this section.
 The interaction between the geometry and the matter is described by the Einstein field equations (we use geometrized units, i.e.
 the gravitational constant G and the speed of light in vacuum c are set equal to one):
\begin{eqnarray*}
G_{\alpha\beta}= 8\pi T_{\alpha \beta}
\end{eqnarray*}
where $G_{\alpha\beta}$ is the Einstein tensor and $T_{\alpha \beta}$ is the energy-momentum tensor. For the matter model
 we will take the point of view of kinetic theory \cite{St}. The sign conventions of \cite{RA} are used. Also the Einstein summation convention
 that repeated indices are to be summed over is used. Latin indices run from one to three and Greek ones from zero to three.

Consider a particle with non-zero rest mass which moves under the influence of the gravitational field. The mean field will be described now by the metric and the components of the metric connection. The wordline $x^\alpha$ of a particle
 is a timelike curve in spacetime. The unit future-pointing tangent vector to this curve is the 4-velocity $v^{\alpha}$ and $p^{\alpha}=mv^{\alpha}$
 is the 4-momentum of the particle. Let $T_x$ be the tangent space at a point $x^{\alpha}$ in the spacetime $M$, then we define the
 \emph{phase space for particles of arbitrary rest masses $P$} to be the following set:
\begin{eqnarray*}
P=\{(x^{\alpha},p^{\alpha}):\ x^{\alpha} \in M,\ p^{\alpha} \in T_x,\ p_{\alpha} p^{\alpha} \leq 0,\ p^{0}>0\}
\end{eqnarray*}
which is a subset of the tangent bundle $TM=\{(x^{\alpha},p^{\alpha}):\ x^{\alpha} \in M,\ p^{\alpha} \in T_x\}$.
For particles of the same type and with the same rest mass $m$ which is given by the mass shell relation:
\begin{eqnarray*}
 p_{\alpha} p^{\alpha}=-m^2
\end{eqnarray*}
we have the \textit{phase space $P_m$ for particles of mass $m$}:
\begin{eqnarray*}
P_m=\{(x^{\alpha},p^{\alpha}):\ x^{\alpha} \in M,\ p^{\alpha} \in T_x,\ p_{\alpha} p^{\alpha}=-m^2,\ p^{0}>0\}
\end{eqnarray*}
We will consider from now on that all the particles have \textit{equal} mass $m$. For how this relates to the general case of 
different masses see \cite{SI}. We will choose units such that $m=1$ which means that a distinction between velocities and momenta is not necessary.
 We have then that the possible values for the 4-momenta are all future pointing unit timelike vectors. These values form the hypersurface:
\begin{eqnarray*}
P_1=\{(x^{\alpha},p^{\alpha}):\ x^{\alpha} \in M,\ p^{\alpha} \in T_x,\ p_{\alpha} p^{\alpha}= -1,\ p^{0}>0\}
\end{eqnarray*}
which we will call the mass shell. The collection of particles (galaxies or clusters of galaxies) will be described (statistically) by a non-negative real valued
 distribution function $f(x^\alpha,p^\alpha)$ on $P_1$. This function represents the density of particles at a given spacetime point with
 given four-momentum. A free particle travels along a geodesic. Consider now a future-directed timelike geodesic parametrized by proper time $s$.
 The tangent vector is then at any time future-pointing unit timelike. Thus the geodesic has a natural lift to a curve on $P_1$ by taking its
 position and tangent vector. The equations of motion thus define a flow on $P_1$ which is generated by a vector field $L$ which is called geodesic
 spray or Liouville operator. The geodesic equations are:
\begin{eqnarray*}
\frac{dx^{\alpha}}{ds}=p^{\alpha}; \ \ \frac{dp^{\alpha}}{ds}=-\Gamma^\alpha_{\beta \gamma} p^{\beta} p^{\gamma}
\end{eqnarray*}
where the components of the metric connection, i.e. 
$\Gamma_{\alpha\beta \gamma}=g(e_{\alpha},\nabla_{\gamma}e_{\beta})=g_{\alpha\delta}\Gamma^{\delta}_{\beta\gamma}$ can be expressed in the
 vector basis $e_\alpha$ as [(1.10.3) of \cite{JS}]:
\begin{align}\label{con}
\Gamma_{\alpha\beta\gamma}=\frac12[e_{\beta}(g_{\alpha\gamma})+e_{\gamma}(g_{\beta\alpha})+e_{\alpha}(g_{\gamma\beta})+\eta^\delta_{\gamma\beta}g_{\alpha\delta}+\eta^{\delta}_{\alpha\gamma}g_{\beta\delta}-\eta^{\delta}_{\beta\alpha}g_{\gamma\delta}]
\end{align}
The commutator of the vectors $e_\alpha$ can be expressed with the following formula:
\begin{eqnarray*}
[e_{\alpha},e_{\beta}]=\eta^{\gamma}_{\alpha \beta}e_{\gamma}
\end{eqnarray*}
where $\eta^{\gamma}_{\alpha \beta}$ are called commutation functions.

 The restriction of the Liouville operator to the mass shell is defined as:
\begin{align*}
L=\frac{dx^{\alpha}}{ds}\frac{\partial}{\partial x^{\alpha}}+\frac{dp^{a}}{ds}\frac{\partial}{\partial p^{a}}.
  \end{align*}
Using the geodesic equations it has the following form
\begin{align*}
 L=p^{\alpha}\frac{\partial}{\partial x^{\alpha}}-\Gamma^a_{\beta \gamma} p^{\beta} p^{\gamma}\frac{\partial}{\partial p^{a}}
\end{align*}
This operator is sometimes also called geodesic spray. If we denote now the phase space density of collisions by $C(f)$,
 then the Boltzmann equation in curved spacetime in our notation looks as follows:
 \begin{align*}
 L(f)=C(f)
 \end{align*}
describes the evolution of the distribution function. Between collisions the particles follow geodesics. We will consider the collisionless
 case which is described via the Vlasov equation:
\begin{align*}
L(f)=0
\end{align*}
\subsection{Energy momentum tensor and characteristics}
The unknowns of our system are a 4-manifold $M$, a Lorentz metric $g_{\alpha\beta}$ on this manifold and the distribution function $f$
 on the mass shell $P_1$ defined by the metric. We have the Vlasov equation defined by the metric for the distribution function and the
 Einstein equations. It remains to define the energy-momentum tensor $T_{\alpha\beta}$ in terms of the distribution and the metric.
Before that we need a Lorentz invariant volume element on the mass shell. A point of a the tangent space has the volume element
$ |g^{(4)}|^{\frac{1}{2}} dp^0 dp^1 dp^2 dp^3$ ($g^{(4)}$ is the determinant of the spacetime metric) which is Lorentz invariant. Now considering $p^0$ as a dependent variable 
the induced (Riemannian) volume of the mass shell considered as a hypersurface in the tangent space at that point is
\begin{eqnarray*}
 \varpi=2H(p^{\alpha})\delta( p_{\alpha} p^{\alpha}+m^2)|g^{(4)}|^{\frac{1}{2}} dp^0 dp^1 dp^2 dp^3
\end{eqnarray*}
where $\delta$ is the Dirac distribution function and $H(p^{\alpha})$ is defined
 to be one if $p^{\alpha}$ is future directed and zero otherwise.
We can write this explicitly as:
 \begin{eqnarray*}
\varpi=|p_0|^{-1} |g^{(4)}|^{\frac{1}{2}} dp^1 dp^2 dp^3
\end{eqnarray*}
Now we define the energy momentum tensor as follows:
 \begin{eqnarray*}
T_{\alpha\beta}=\int f(x^{\alpha},p^{a}) p_{\alpha}p_{\beta}\varpi
\end{eqnarray*}
One can show that $T_{\alpha\beta}$ is divergence-free and thus it is compatible with the Einstein equations. For collisionless
 matter all the energy conditions hold (for details we refer to \cite{RIV}).  In particular the dominant energy condition is equivalent to the statement that
 in any orthonormal basis the energy density dominates the other components of $T_{\alpha\beta}$, i.e. $T_{\alpha\beta}\leq T_{00}$
 for each $\alpha,\beta$ (P. 91 of \cite{HaEl}). Using the mass shell relation one can see that this holds for collisionless matter.
 The non-negative sum pressures condition is in our case equivalent to $g_{ab}T^{ab} \ge 0$.

The Vlasov equation in a fixed spacetime can be solved by the method of characteristics:
\begin{eqnarray*}
\frac{dX^{a}}{ds}=P^{a}; \ \ \frac{dP^{a}}{ds}=-\Gamma^a_{\beta \gamma} P^{\beta} P^{\gamma}
\end{eqnarray*}
Let $X^a(s,x^{\alpha},p^a)$, $P^a(s,x^{\alpha},p^a)$ be the unique solution of that equation with initial conditions
 $X^a(t,x^{\alpha},p^a)=x^a$ and $P^a(t,x^{\alpha},p^a)=p^a$. Then the solution of the Vlasov equation can be written as:
\begin{eqnarray*}
f(x^{\alpha},p^a)=f_0(X^a(0,x^{\alpha},p^a),P^a(0,x^{\alpha},p^a))
\end{eqnarray*}
where $f_0$ is the restriction of $f$ to the hypersurface $t=0$. It follows that if $f_0$ is bounded the same is true for $f$. 
We will assume that $f$ has compact support in momentum space for each fixed t. This property
holds if the initial datum $f_0$ has compact support and if each hypersurface $t = t_0$ is a Cauchy hypersurface. Note that there is no obvious
 reason why a solution of the Boltzmann equation with compactly supported initial data should have compact support \cite{RIV}.
\subsection{The initial value problem}

Before coming to our symmetry assumption we want to briefly introduce the initial value problem for the Einstein-Vlasov system.
 For a general introduction to the initial value problem in general relativity we refer to \cite{HR} and for the Einstein-Vlasov system
 in particular we refer to \cite{RIV}. In general the initial data for the Einstein-matter equations consist of a metric $g_{ab}$ on the
 initial hypersurface, the second fundamental form $k_{ab}$ on that hypersurface and some matter data. Thus we have a Riemannian metric $g_{ab}$,
 a symmetric tensor $k_{ab}$ and some matter fields defined on an abstract 3-dimensional manifold $S$.

Solving the initial value problem means embedding $S$ into a 4-dimensional $M$ on which are defined a Lorentzian metric $g_{\alpha\beta}$
 and matter fields such that $g_{ab}$ and $k_{ab}$ are the pullbacks to $S$ of the induced metric and second fundamental form of the image of
 the embedding of $S$ while $f$ is the pullback of the matter fields. Finally $g^{\alpha\beta}$ and $f$ have to satisfy the Einstein-matter
 equations.

For the Einstein-Vlasov system it has been shown \cite{CBr} that given an initial data set there exists a corresponding solution
 of the Einstein-Vlasov system and that this solution is locally unique up to diffeomorphism (see also theorem 1.1 of \cite{RIV}). The extension
 to a global theorem has not been achieved yet. However if one assumes that the initial data have certain symmetry, this symmetry is inherited 
by the corresponding solutions (see 5.6 of \cite{HFAR} for a discussion). In particular for the case we will deal with, i.e. expanding Bianchi
 models (except type IX) coupled to dust or to collisionless matter the spacetime is future complete (theorem 2.1 of \cite{GC}).

\section{Bianchi spacetimes}
\subsection{Definition of Bianchi spacetimes}
We start with the definition of homogeneity of spacetimes taken from chapter 5.1 of \cite{WAbook}. 
\begin{deff}
A spacetime $(M, g_{\alpha\beta})$ is said to be (spatially) \textit{homogeneous} if there exists a one-parameter family of spacelike hypersurfaces
 $S_{t}$ foliating the spacetime such that for each t and for any points $P,Q \in S_{t}$ there exists an isometry of the spacetime metric,
 $g_{\alpha\beta}$, which takes $P$ into $Q$.
\end{deff}

The basis for the classification of homogeneous spacetimes is the work of Bianchi \cite{BI}
 which was introduced to cosmology by Taub \cite{Taub}. Here we will use the modern terminology and we define Bianchi spacetimes as follows:
\begin{deff}
A \textit{Bianchi spacetime} is defined to be a spatially homogeneous spacetime whose isometry group possesses a three-dimensional subgroup $G$
 that acts simply transitively on the spacelike orbits.
\end{deff}
Not all homogeneous spacetimes are Bianchi spacetimes. But the only case where $G$ does not act simply transitively or does not possess
 a subgroup with simply transitive action are the so called Kantowski-Sachs models. The Bianchi models can be subclassified into two classes
 \cite{EM}:
 class A and B. Later we will only deal with Bianchi class A, however all the equations in this s are valid for Bianchi spacetimes in general.

The only Bianchi spacetimes which admit a compact Cauchy hypersurface are Bianchi I and IX. In order to be not that restrictive we will
 consider locally spatially homogeneous spacetimes. They are defined as follows. Consider an initial data set on a three-dimensional manifold $S$. 
Then this initial data set is called locally spatially homogeneous if the naturally associated data set on the universal covering $\tilde{S}$ is
 homogeneous. For Bianchi models the universal covering space $\tilde{S}$ can be identified with its Lie group $G$ (see \cite{CC}, \cite{GC} for details).
 
\subsection{Description of Bianchi spacetimes via the metric approach}

A Bianchi spacetime admits a Lie algebra of Killing vector fields with basis $\mathbf{k}_1$, $\mathbf{k}_2$, $\mathbf{k}_3$ and
 structure constants $C^{c}_{ab}$, such that:
\begin{eqnarray*}
[\mathbf{k}_a,\mathbf{k}_b]=-C^{c}_{ab}\mathbf{k}_c.
\end{eqnarray*}
The Killing vector fields $\mathbf{k}_a$ are tangent to the group orbits which are called surfaces of homogeneity. If one chooses a unit
 vector field $n$ normal to the group orbits we have a natural choice for the time coordinate $t$ such that the group orbits are given by
 a constant $t$. This unit normal is invariant under the group, i.e:
\begin{eqnarray*}
 [\mathbf{n},\mathbf{k}_a]=0
\end{eqnarray*}
One can now choose a triad of spacelike vectors $e_{a}$ that are tangent to the group orbits:
\begin{eqnarray*}
g(\mathbf{n},\mathbf{e}_a)=0
\end{eqnarray*}
and that commute with the Killing vector fields:
\begin{eqnarray*}
[\mathbf{e}_a,\mathbf{k}_b]=0
\end{eqnarray*}
A frame $\{\mathbf{n},\mathbf{e}_{a}\}$ chosen in this way is called a left invariant frame and it is generated by the right invariant
 Killing vector fields. Since $\mathbf{n}$ is hypersurface orthogonal the vector fields
 $\mathbf{e}_a$ generate a Lie algebra with structure constants $\eta^c_{ab}$. It can be shown that this Lie algebra is in fact equivalent
 to the Lie algebra of the Killing vector fields. Thus one can classify the Bianchi spacetimes using either the structure constants or the spatial
 commutation functions of the basis vectors. The remaining freedom in the choice of the frame is a time-dependent linear transformation,
 which can be used to introduce a set of time-independent spatial vectors $\mathbf{E}_a$:
\begin{eqnarray*}
 [\mathbf{E}_a,\mathbf{n}]=0.
\end{eqnarray*}
The corresponding commutation functions are then constant in time and one can make them equal to the structure constants:
\begin{eqnarray*}
[\mathbf{E}_a,\mathbf{E}_b]=C^c_{ab}\mathbf{E}_c
\end{eqnarray*}
This is our choice which is sometimes called the metric approach. If $\mathbf{W}^a$ denote the 1-forms dual to the frame vectors $\mathbf{E}_a$
the metric of a Bianchi spacetime takes the form:
\begin{eqnarray}\label{mt}
^4 g=-dt^2+g_{ab}(t)\mathbf{W}^a \mathbf{W}^b
\end{eqnarray}
where $g_{ab}$ (and all other tensors) on $G$ will be described in terms of the frame components of the left invariant frame which has been
 introduced.
A dot above a letter will denote a derivative with respect to the cosmological time $t$.

\subsection{3+1 Decomposition of the Einstein equations}

We will use the 3+1 decomposition of the Einstein equations as made in \cite{RA}. Comparing our metric (\ref{mt}) with (2.28) of \cite{RA}
 we have that $\alpha=1$ and $\beta^a=0$ which means that the lapse function is the identity and the shift vector vanishes. There the abstract index 
notation is used. We can interpret the quantities as being frame components. For details we refer to chapter 2.3 of \cite{RA}. There are different
 projections of the energy momentum tensor which are important 
\begin{eqnarray*}
   \rho&=&T^{00}\\
     j_a&=&T_a^0\\
  S_{ab}&=&T_{ab}
\end{eqnarray*}
where $\rho$ is the energy density and $j_a$ is the matter current.

The second fundamental form $k_{ab}$ can be expressed as:
\begin{eqnarray}\label{a}
 \dot{g}_{ab}=-2 k_{ab}.
\end{eqnarray}
The Einstein equations:
\begin{eqnarray}\label{EE}
 \dot{k}_{ab}=R_{ab}+k~k_{ab}-2k_{ac}k^c_b-8\pi(S_{ab}-\frac{1}{2}g_{ab}S)-4\pi\rho g_{ab}
\end{eqnarray}
where we have used the notations $S =g^{ab}S_{ab}$, $k = g^{ab}k_{ab}$, and $R_{ab}$ is the Ricci tensor of the three-dimensional metric.
The evolution equation for the mixed version of the second fundamental form is (2.35) of \cite{RA}:
\begin{eqnarray}\label{MV}
 {{\dot{k}}^a}_{b}=R^a_b+k~k^a_b - 8 \pi S^a_b + 4 \pi \delta^a_b (S -\rho)
\end{eqnarray}
From the constraint equations since $k$ only depends on the time variable we have that:
\begin{eqnarray}\label{CE1}
 R-k_{ab}k^{ab}+k^2&=&16\pi\rho\\
\label{CE2}
 \nabla^a k_{ab}& = & 8 \pi j_b
\end{eqnarray}
where $R$ is the Ricci scalar curvature.

Another useful relation concerns the determinant $g$ of the induced metric ((2.30) of \cite{RA}):
\begin{eqnarray}\label{det}
\frac{d}{dt}(\log g)=-2 k
\end{eqnarray}
Taking the trace of (\ref{MV}):
\begin{eqnarray}\label{im}
 \dot k=R+k^2+4\pi  S -12\pi \rho
\end{eqnarray}
With (\ref{CE1}) one can eliminate the energy density and (\ref{im}) reads:
\begin{eqnarray}\label{in}
 \dot k=\frac{1}{4}(k^2+R+3k_{ab}k^{ab})+4\pi S
\end{eqnarray}
Finally if one substitutes for the Ricci scalar with (\ref{CE1}):
\begin{eqnarray}\label{loo}
 \dot k=k_{ab}k^{ab}+4\pi (S+\rho)
\end{eqnarray}
\subsection{Time origin choice and new variables}\label{tc}

Now with the 3+1 formulation our initial data are $(g_{ij}(t_0), k_{ij}(t_0),f(t_0))$, i.e. a Riemannian metric, a second fundamental
 form and the distribution function of the Vlasov equation, respectively, on a three-dimensional manifold $S(t_0)$. This is the initial data set
 at $t=t_0$ for the Einstein-Vlasov system.

We assume that $k < 0$ for all time following \cite{CC} (see comments below lemma 2.2 of \cite{CC}). This enables us given $k(t_0)$ to set without
 loss of generality $t_0=-2/k(t_0)$, since future geodesic completeness is known and one can always add an arbitrary constant to the
time origin. The reason for this choice will become clear later and is of technical nature.

We will now introduce several new variables in order to use the ones which are common in Bianchi cosmologies and to be able
 to compare results. We can decompose the second fundamental form introducing $\sigma_{ab}$ as the trace-free part:
\begin{eqnarray}\label{TF}
 k_{ab}=\sigma_{ab}-H g_{ab}
\end{eqnarray}
\begin{eqnarray}\label{tf}
 k_{ab}k^{ab}=\sigma_{ab}\sigma^{ab}+3H^2
\end{eqnarray}
Using the Hubble parameter:
\begin{eqnarray*}
H=-\frac{1}{3}k
\end{eqnarray*}
we define:
\begin{eqnarray}\label{ab}
\Sigma_a^b=\frac{\sigma_a^b}{H}
\end{eqnarray}
and
\begin{eqnarray}
&&\Sigma_{+}=-\frac12(\Sigma_{2}^2+\Sigma_{3}^3) 
\\
&& \Sigma_{-}=-\frac{1}{2\sqrt{3}}(\Sigma_{2}^2-\Sigma_{3}^3) 
\end{eqnarray}
Define also:
\begin{eqnarray}
\Omega=8\pi \rho/3H^2\\
\label{q} q=-1-\frac{\dot{H}}{H^2}\\
\label{tau} \frac{d\tau}{dt}=H
\end{eqnarray}
The time variable $\tau$ is dimensionless and sometimes very useful. From (\ref{CE1}) we obtain the constraint equation:
\begin{eqnarray*}
 \frac{1}{6H^2}(R-\sigma_{ab}\sigma^{ab})=\Omega-1
\end{eqnarray*}
and from (\ref{in}) the evolution equation for the Hubble variable:
\begin{eqnarray}\label{H-1}
\partial_t(H^{-1})=\frac{3}{2}+\frac{1}{12}(\frac{R}{H^2}+\frac{3}{H^2}\sigma_{ab}\sigma^{ab})+\frac{4\pi S}{3H^2}
\end{eqnarray}
Combining the last two equations with (\ref{MV}) we obtain the evolution equations for $\Sigma_-$ and $\Sigma_+$:
\begin{eqnarray}
\label{Bla1}&&\dot{\Sigma}_+=H[\frac{2R-3(R^2_2+R^3_3)}{6H^2}-\Sigma_+(3+\frac{\dot{H}}{H^2})+\frac{4\pi}{3H^2}(3S^2_2+3S^3_3-2S)]\\
\label{Bla2}&&\dot{\Sigma}_-=H[\frac{R_3^3-R^2_2}{2\sqrt{3}H^2}-(3+\frac{\dot{H}}{H^2})\Sigma_-+\frac{4\pi(S^2_2-S^3_3)}{\sqrt{3}H^2}]
\end{eqnarray}
\subsection{Vlasov equation with Bianchi symmetry}

Since we use a left-invariant frame $f$ will not depend on $x^a$ and the Vlasov equation takes the form:
\begin{eqnarray*}
p^0\frac{\partial f}{\partial t}-\Gamma^a_{\beta\gamma}p^{\beta}p^{\gamma}\frac{\partial f}{\partial p^a}=0
\end{eqnarray*}
It turns out that the equation simplifies if we express $f$ in terms of $p_i$ instead of $p^{i}$ what we can do due to the mass shell
 relation:
\begin{eqnarray*}
p^0\frac{\partial f}{\partial t}-\Gamma_{a\beta\gamma}p^{\beta}p^{\gamma}\frac{\partial f}{\partial p_a}=0
\end{eqnarray*}
Because of our special choice of frame the metric has the simple form (\ref{mt}). This has the consequence that only the spatial
 components of the metric connection remain and that the first three terms of (\ref{con}) vanish. Due to the fact that we are contracting and
 the antisymmetry of the structure constant we finally arrive at:
\begin{eqnarray}\label{ve}
 \frac{\partial f}{\partial t}+(p^0)^{-1}C^d_{ba}p^{b}p_{d}\frac{\partial f}{\partial p_a}=0
\end{eqnarray}
From (\ref{ve}) it is also possible to define the characteristic curve $V_a$:
\begin{eqnarray}\label{charak}
 \frac{dV_a}{dt}=(V^0)^{-1}C^d_{ba}V^bV_{d}
\end{eqnarray}
for each $V_i(\bar{t})=\bar{v}_i$ given $\bar{t}$. Note that if we define:
\begin{eqnarray}\label{VVV}
 V=g^{ij}V_iV_j
\end{eqnarray}
due to the antisymmetry of the structure constants we have with (\ref{charak}):
\begin{eqnarray}\label{ha}
\frac{dV}{dt}=\frac{d}{dt}(g^{ij})V_iV_j
\end{eqnarray}
Let us also write down the components of the energy momentum tensor in our frame:
\begin{eqnarray}
&&T_{00}=\int f(t,p^{a}) p^0 \sqrt{g}dp^1 dp^2 dp^3\\
\label{emt2}&&T_{0j}=-\int f(t,p^{a}) p_j \sqrt{g}dp^1 dp^2 dp^3\\
\label{emt3}&&T_{ij}=\int f(t,p^{a}) p_i p_j (p^0)^{-1}\sqrt{g}dp^1 dp^2 dp^3
\end{eqnarray}

\section{Bianchi A spacetimes}
\subsection{Definition and classification of Bianchi A spacetimes}
In the last chapter we have presented the Einstein-Vlasov system with Bianchi symmetry. However our results concern only a special class
 of the Bianchi spacetimes, namely that of class $A$.
\begin{deff}
A \textit{Bianchi A spacetime} is a Bianchi spacetime whose three-dimensional Lie algebra has traceless structure constants, i.e. $C^a_{ba}=0$.
\end{deff}
In that case there is a unique symmetric matrix, called commutator matrix with components $\nu^{ij}$ such that the structure constants
 can be written as follows (lemma 19.3 of \cite{HR}):
\begin{eqnarray}\label{sc}
C^{a}_{bc}=\varepsilon_{bcd}\nu^{da}
\end{eqnarray}

The transformation rule of the commutator matrix under a change of basis of the Lie algebra can be used to classify the Bianchi class A
 Lie algebras. It is possible to diagonalize $\nu$ and the diagonal elements $\nu_1$, $\nu_2$ and $\nu_3$ can be used to classify the different
 Bianchi types of class A (for details see chapter 19.1 of \cite{HR}). We will study Bianchi II and VI$_0$. For Bianchi II we have that the only
 non-vanishing element of $\nu$ is $\nu_1=1$, thus from (\ref{sc}) we see that for Bianchi II the only non-vanishing structure constants are:
\begin{eqnarray}\label{sc2}
C^1_{23}=1=-C^1_{32}
\end{eqnarray}
In the case of Bianchi VI$_0$ the only non-vanishing elements of $\nu$ are $\nu_2=1$ and $\nu_3=-1$ thus the only non-vanishing
 structure constants are:
\begin{eqnarray}\label{sc6}
C^2_{31}=1=-C^2_{13},\ \ C^3_{21}=1=-C^3_{12}
\end{eqnarray}

\subsection{Reflection symmetry and vanishing tilt}
We will assume an additional symmetry namely the reflection symmetry such that the matter current vanishes. This will be a restriction to the diagonal case.
 The reflection symmetry has been defined in (2.10) of \cite{IS} for the case of
 Bianchi I, but one can define this for other Bianchi types as well with the difference that the distribution function now will depend in general
 on the time variable:
\begin{eqnarray*}
 f(t,p_1,p_2,p_3)=f(t,-p_1,-p_2,p_3)=f(t,p_1,-p_2,-p_3)
\end{eqnarray*}
Suppose that the distribution function is initially reflection symmetric and the metric and the second fundamental form are initially diagonal. 
Then from (\ref{emt2})-(\ref{emt3}) we see that the energy-momentum tensor is diagonal as well. 
From (\ref{a}) and (\ref{MV}) we can see that the metric and the second fundamental will remain diagonal. This symmetry implies in particular that there is no matter current, which means that there is no 'tilt'.
Initial data with symmetry lead to solutions of the Einstein-Vlasov equations with symmetry, for a proof of this fact we refer to 9.2-9.3 of \cite{RA}.
\subsection{Some formulas for the diagonal case}
Now we will introduce formulas which are valid in the diagonal case. There $(ijk)$ denotes a cyclic permutation of $(123)$ and the
 Einstein summation convention is suspended for the first two formulas. Let us define:
\begin{eqnarray*}
n_i=\nu_i \sqrt{\frac{g_{ii}}{g_{jj}g_{kk}}}
\end{eqnarray*}
The Ricci tensor is given by (11a) of \cite{CH}:
\begin{eqnarray*}
R^i_i=\frac{1}{2}[n_i^2 -(n_j-n_k)^2]
\end{eqnarray*}
We define:
\begin{eqnarray*}\label{444}
 N_i=\frac{n_i}{H}
\end{eqnarray*}
In the diagonal case we have:
\begin{eqnarray*}
\Sigma_+^2+\Sigma_-^2=\frac16 \frac{\sigma_{ab}\sigma^{ab}}{H^2}
\end{eqnarray*}
from which follows that the constraint equation can be written in the following way:
\begin{eqnarray}\label{co}
\Sigma_+^2+\Sigma_-^2=\Omega-1-\frac{1}{6H^2}R
\end{eqnarray}
Now we proceed to use them for the cases of Bianchi II and VI$_0$.
\subsubsection{Expressions for diagonal Bianchi II}
For Bianchi II we have then:
\begin{eqnarray*}
 R_{1}^1=-R_{2}^2=-R_{3}^3=-R=\frac{1}{2} n_1^2
\end{eqnarray*}
From the constraint equation (\ref{co}) we obtain:
\begin{eqnarray*}
\Sigma_+^2+\Sigma_-^2=1-\Omega-\frac{1}{12}N_1^2
\end{eqnarray*}
and from (\ref{H-1}) we obtain the equation for the evolution of $H$:
\begin{eqnarray}\label{1}
\partial_{t}(H^{-1})=\frac{3}{2}-\frac{N_1^2}{24}+\frac{3}{2}(\Sigma_+^2+\Sigma_-^2)+\frac{4\pi S}{3H^2}
\end{eqnarray}
and the evolution equation for $\Sigma_+$ and $\Sigma_-$:
\begin{eqnarray}
\label{2}&&\dot{\Sigma}_+=H[ \frac{1}{3}N_1^2-(3+\frac{\dot{H}}{H^2})\Sigma_++\frac{4\pi}{3H^2}(S^2_2+S^3_3-2S^1_1)]\\
\label{3}&&\dot{\Sigma}_-=H[-(3+\frac{\dot{H}}{H^2})\Sigma_-+\frac{4\pi}{\sqrt{3}H^2}(S^2_2-S^3_3)]
\end{eqnarray}
From the definition (\ref{a}) for the second fundamental form the evolution equation for $n_1^2$ follows:
\begin{eqnarray*}
 \frac{d}{dt}({n}^2_1)=2(-4\sigma_+-H)n_1^2
\end{eqnarray*}
In terms of $N_1^2$:
\begin{eqnarray*}
 \frac{d}{dt}({N}_1^2)=-2N_1^2H(4\Sigma_++1+\frac{\dot{H}}{H^2})
\end{eqnarray*}
or
\begin{eqnarray}\label{4}
 \dot{N}_1&=&-N_1H(4\Sigma_++1+\frac{\dot{H}}{H^2})
\end{eqnarray}
Let us write the equations (\ref{2})-(\ref{4}) with $\tau$ and $q$:
\begin{eqnarray}
\label{01}&&\Sigma_+'=\frac{1}{3}N_1^2-(2-q)\Sigma_++\frac{4\pi}{3H^2}(S^2_2+S^3_3-2S^1_1)\\
\label{22}&&\Sigma_-'=-(2-q)\Sigma_-+\frac{4\pi}{\sqrt{3}H^2}(S^2_2-S^3_3)\\
\label{33}&& N_1'=N_1(q-4\Sigma_+)
\end{eqnarray}

Note that these equations are the same as (6.21) of \cite{JW} with $\gamma=1$ if one sets $S=0$ in (\ref{01})-(\ref{33}).

\subsubsection{Expressions for diagonal Bianchi VI$_0$}
For Bianchi VI$_0$ we have then:
\begin{eqnarray*}
&& R_{1}^1=R=-\frac{1}{2} (n_2-n_3)^2\\
&&R^2_2=-R^3_3=\frac{1}{2}(n_2^2-n_3^2)
\end{eqnarray*}
The constraint equation (\ref{co}) is:
\begin{eqnarray*}
\Sigma_+^2+\Sigma_-^2=1-\Omega-\frac{1}{12}(N_2-N_3)^2
\end{eqnarray*}
and the evolution equations
\begin{eqnarray*}
&&\partial_{t}(H^{-1})=\frac{3}{2}-\frac{1}{24}(N_2-N_3)^2+\frac{3}{2}(\Sigma_+^2+\Sigma_-^2)+\frac{4\pi S}{3H^2}\\
&&\dot{\Sigma}_+=H[-\frac{1}{6}(N_2-N_3)^2-\Sigma_+(3+\frac{\dot{H}}{H^2})+\frac{4\pi}{3H^2}(S^2_2+S^3_3-2S^1_1)]\\
&&\dot{\Sigma}_-=H[\frac{N_3^2-N_2^2}{2\sqrt{3}}-(3+\frac{\dot{H}}{H^2})\Sigma_-+\frac{4\pi(S^2_2-S^3_3)}{\sqrt{3}H^2}]\\
&&\dot{N}_2=-N_2H(-2\Sigma_+-2\sqrt{3}\Sigma_-+1+\frac{\dot{H}}{H^2})\\
&&\dot{N}_3=-N_3H(-2\Sigma_++2\sqrt{3}\Sigma_-+1+\frac{\dot{H}}{H^2})
\end{eqnarray*}

In analogy to Bianchi II these equations can be compared to (6.9)-(6.10) of \cite{JW} setting $N_1$ to zero and using the definition of $q$ (\ref{q}).

\section{Special solutions}
In this section we present some special solutions which will be important, since we will show that the late time asymptotics of the 
Bianchi types considered behave like them in a sense which will be specified later. We start with the Kasner solution which is the general Bianchi I 
vacuum solution to motivate also the concept of generalized Kasner exponents.
\subsection{The Kasner solution and generalized Kasner exponents}
The Kasner solution \cite{KA} is the general Bianchi I vacuum solution, thus all components of the energy-momentum tensor and the scalar
 curvature $R$ vanish. From the constraint equation one obtains:
\begin{eqnarray*}
 \Sigma_+^2+\Sigma_-^2=1
\end{eqnarray*}
which is known as the Kasner circle. The metric components are:
\begin{eqnarray*}
 g_{ij}=\diag (t^{2p_1}, t^{2p_2}, t^{2p_3})
\end{eqnarray*}
where $p_1$, $p_2$ and $p_3$ satisfy:
\begin{eqnarray*}
 p_1+p_2+p_3=1 \\
p_1^2+p_2^2+p_3^2=1
\end{eqnarray*}
One can easily compute that the Hubble variable is $H=\frac{1}{3}t^{-1}$.

For more general spacetimes let $\lambda_i$ be the eigenvalues of  $k_{ij}$ with respect to $g_{ij}$, i.e., the solutions of:
\begin{eqnarray}\label{ev}
\det (k^i_j-\lambda \delta^i_j)=0 
\end{eqnarray}
 We define
\begin{eqnarray*}
p_i=\frac{\lambda_i}{k} 
\end{eqnarray*}
as the \emph{generalized Kasner exponents}. They satisfy the first but in general not the second Kasner relation.
\subsection{The Collins-Stewart solution}
Another special solution which will play an important role is the Collins-Stewart solution (\cite{CS}, p. 430) with dust ($\gamma=1$) 
which has Bianchi II symmetry:
\begin{eqnarray*}
 g_{CS}=\diag(2t, (2t)^{3/2}, (2t)^{3/2})
\end{eqnarray*}
The Hubble parameter is $H=\frac{2}{3}t^{-1}$ and the energy density $8\pi\rho_{CS}=\frac{5}{4}t^{-2}$.
The values of the variables which have been introduced previously are:
\begin{eqnarray*}
\Sigma_{+}=\frac{1}{8}; \ \Sigma_{-}=0; \ \Omega=\frac{15}{16}; \ N_1=\frac{3}{4}
\end{eqnarray*}
\subsection{The Ellis-MacCallum solution}
In the case of Bianchi VI$_0$ there is a dust solution with diagonal metric discovered by Ellis and MacCallum (\cite{EM}, pp. 124-125):
\begin{eqnarray*}
 g_{EM}=\diag (t^2, t^1 ,t^1)
\end{eqnarray*}
The Hubble parameter is $H=\frac{2}{3}t^{-1}$ as in the Collins-Stewart solution, but the 
energy density $8\pi\rho_{EM}=t^{-2}$ is different. Here the values of the introduced variables are:
\begin{eqnarray*}
\Sigma_+= -\frac{1}{4}; \ \Sigma_-=0; \ N_1=0; \ N_2=-N_3=\frac{3}{4}; \ \Omega=\frac{3}{4}
\end{eqnarray*}
The generalization of this solution to different values of $\gamma$ is called Collins solution.

\section{Einstein-dust with small data}\label{peta}
\subsection{Einstein-dust system}
Let us present briefly the Einstein-Euler system. We will consider the isentropic case where the matter fields are the energy density
 $\rho$ and the four-velocity $u^{\alpha}$ which is a unit timelike vector. The equation of state $P=f(\rho)$ relates the pressure $P$ with
 the energy density. The energy-momentum tensor is:
\begin{eqnarray*}
T_{\alpha\beta}=(\rho+P)u_{\alpha}u_{\beta}+P g_{\alpha\beta}
\end{eqnarray*}
and the equations of motion are equivalent to the condition that the energy-momentum tensor is divergence-free.
The Einstein-dust system is then obtained via the condition $P=0$. It can be seen as a very singular solution of the Einstein-Vlasov system.
 Formally the system can be obtained from the Einstein-Vlasov system choosing $f$ to be of the form:
\begin{eqnarray*}
f(t,x^a,p^a)=|u_0||g^{(4)}|^{-\frac12}\rho(t,x^a)\delta(p^a-u^a)
\end{eqnarray*}
where $u_0$ is obtained via the mass shell relation. The relation of the Einstein-dust system to the Einstein-Vlasov system is in general
 subtle and we refer to \cite{Rendall} for more information on that. Here we will look at the special solutions of the corresponding
 Einstein-dust systems in order to obtain some intuition about the Einstein-Vlasov system. It is easy to see that the special solutions
 are equilibrium points. The stability of these equilibrium points has already been studied (see for instance \cite{JW}). For the case of
 Bianchi I there even exist a general expression [(11-1.12) of \cite{HS}], where one can see that isotropization occurs. 
Actually for all the Bianchi cases we study here, Liapunov functions have been found, such that besides the stability
 also the global behaviour is known. In this section we will start dealing with estimates. $C$ will denote an arbitrary constant
 and $\epsilon$ a small and strictly positive constant. They both may appear several times in different equations or inequalities
 without being the same constant. 

\subsection{Linearization of Einstein-dust around Collins-Stewart}
Let us look at the stability of the Collins-Stewart solution with dust. For the Collins-Stewart solution we have
 $(\Sigma_+=\frac{1}{8},\Sigma_-=0, N_1=\frac{3}{4})$ which is an equilibrium point of the system (\ref{01})-(\ref{33}) with $S=0$.
 Let us translate the equilibrium point to the origin by introducing the variables $\tilde{\Sigma}_+=\Sigma_+-\frac{1}{8}$, 
$\tilde{\Sigma}_-=\Sigma_-$ and $\tilde{N_1}=N_1-\frac{3}{4}$. The linearization is:
\begin{eqnarray*}
 \left(
\begin{matrix} \tilde{\Sigma}_+ \\
\tilde{\Sigma}_- \\
\tilde{N_1}
\end{matrix} \right)'=\left(
\begin{matrix} -\frac{93}{64} & 0 & \frac{63}{128}  \\
0 & -\frac{3}{2} & 0 \\
-\frac{87}{32} & 0 & -\frac{3}{64}
\end{matrix} \right)\left(
\begin{matrix} \tilde{\Sigma}_+ \\
\tilde{\Sigma}_- \\
\tilde{N_1}
\end{matrix} \right)
\end{eqnarray*}
The variable $\Sigma_-$ decouples and we obtain:
 \begin{eqnarray*}
 \Sigma_-=\Sigma_-(\tau_0)e^{-\frac{3}{2}(\tau-\tau_0)}
\end{eqnarray*}
The rest of the system:
\begin{eqnarray}\label{II}
 \left(
\begin{matrix} \tilde{\Sigma}_+ \\
\tilde{N_1}
\end{matrix} \right)'=\left(
\begin{matrix} -93/64 & 63/128  \\
-87/32 & -3/64
\end{matrix} \right)\left(
\begin{matrix} \tilde{\Sigma}_+ \\
\tilde{N_1}
\end{matrix} \right)
\end{eqnarray}
has eigenvalues
\begin{eqnarray*}
 \lambda_{1/2}&=&-\frac{3}{4}(1 \mp  i\sqrt{\frac{3}{2}})\\
\end{eqnarray*}

Translated to our time variable means that the expected estimates for the Vlasov case are:
\begin{eqnarray*}
\Sigma_+-\frac{1}{8}=O(t^{-\frac{1}{2}})\\
N_1-\frac{3}{4}=O(t^{-\frac{1}{2}})\\
\Sigma_-=O(t^{-1})
\end{eqnarray*}
Whether this is true we do not know at this point. We will start proving estimates in the next chapter. 
Here we have obtained these estimates just in order to get a hint about the non-linear behaviour. 
For instance it could be the case that the variable $\Sigma_-$ has the same decay as the other variables. 
However these estimate will turn out to be not sufficient. In addition to assume that we are close to the special solutions,
 we will have to assume that we are close to the dust case. This is done via a momentum bound. 
We have a number (different from zero) of particles at possibly different momenta
 and we define $P$ as the supremum of the absolute value of these momenta at a given time $t$:
\begin{eqnarray*}
 P(t)=\sup \{ \vert p \vert =(g^{ab}p_a p_b)^\frac{1}{2} \vert f(t,p)\neq 0\}
\end{eqnarray*}
A bound on that quantity can be used for estimates on $S/H^2$ as we show now. Consider an orthonormal frame and denote the components
 of the spatial part of the energy-momentum tensor in this frame by $\widehat{S}_{ab}$. The components can be bounded by
\begin{eqnarray*}
\widehat{S}_{ab} \leq P^2(t) \rho
\end{eqnarray*}
so we have that
\begin{eqnarray}\label{PPP}
\frac{\widehat{S}}{\rho} \leq 3P^2 
\end{eqnarray}
Since the Ricci scalar is non-positive it follows from (\ref{CE1}) using the trace-free part of the second fundamental form that:
\begin{eqnarray*}
16\pi \rho =6H^2 + R -\sigma_{ab}\sigma^{ab}
\end{eqnarray*}
Thus we obtain:
\begin{eqnarray*}
16\pi \rho \leq 6H^2
\end{eqnarray*}
Thus from \ref{PPP} we obtain that:
\begin{eqnarray*}
\frac{4\pi S}{3H^2}\leq \frac{3}{2}P^2.
\end{eqnarray*}
For $V$ due to (\ref{VVV})
 and introducing the Collins-Stewart solution we have:
\begin{eqnarray*}
\dot{V}=-t^{-1}g^{11}V_1^2-\frac32t^{-1}(g^{22}V_2^2+g^{33}V_3^2)
\end{eqnarray*}
We see that $\dot{V}\leq -t^{-1}V$ which implies that the following holds:
\begin{eqnarray*}
P=O(t^{-\frac12})
\end{eqnarray*}
\subsection{Linearization of Einstein-dust around Ellis-MacCallum }
For $S=0$ we have:
\begin{eqnarray}
&&\label{A}\Sigma_+'=-\frac{1}{6}(N_2-N_3)^2-\Sigma_+(2-q)\\
&&\label{B}\Sigma_-'=\frac{N_3^2-N_2^2}{2\sqrt{3}}-(2-q)\Sigma_-\\
&&\label{C}N_2'=N_2(2\Sigma_++2\sqrt{3}\Sigma_-+q)\\
&&\label{E}N_3'=N_3(2\Sigma_+-2\sqrt{3}\Sigma_-+q)
\end{eqnarray}
with $q=\frac{1}{2}-\frac{1}{24}(N_2-N_3)^2+\frac{3}{2}(\Sigma_+^2+\Sigma_-^2)$.

For the Ellis-MacCallum solution we have $(\Sigma_+=-\frac{1}{4},\Sigma_-=0, N_2=-N_3=\frac{3}{4})$ which is an equilibrium point of the system
 (\ref{A})-(\ref{E}). Introducing $\tilde{\Sigma}_+=\Sigma_++\frac{1}{4}$, $\tilde{\Sigma}_-=\Sigma_-$, $\tilde{N_2}=N_2-\frac{3}{4}$ and
$\tilde{N_3}=N_3+\frac{3}{4}$. The linearization is:
\begin{eqnarray}\label{VI0}
 \left(
\begin{matrix} \tilde{\Sigma}_+ \\
\tilde{\Sigma}_- \\
\tilde{N_2}\\
\tilde{N_3}
\end{matrix} \right)'=\left(
\begin{matrix} -\frac{21}{16} & 0 & -\frac{15}{32}& \frac{15}{32} \\
0 & -\frac{3}{2} & -\frac{\sqrt{3}}{4} &-\frac{\sqrt{3}}{4} \\
\frac{15}{16} & \frac{3}{2}\sqrt{3} & -\frac{3}{32} & \frac{3}{32} \\
-\frac{15}{16} & \frac{3}{2}\sqrt{3} & \frac{3}{32} & -\frac{3}{32}
\end{matrix} \right)\left(
\begin{matrix} \tilde{\Sigma}_+ \\
\tilde{\Sigma}_- \\
\tilde{N_2}\\
\tilde{N_3}
\end{matrix} \right)
\end{eqnarray}
The eigenvalues are:
\begin{eqnarray*}
 \lambda_{1/2}&=&-\frac{3}{4} (1\pm  i\sqrt{3})\\
\lambda_{3/4}&=&-\frac{3}{4} (1\pm  i)
\end{eqnarray*}
Translated to our time variable the expected estimates for the Vlasov case are:
\begin{eqnarray*}
\Sigma_++\frac{1}{4}=O(t^{-\frac{1}{2}})\\
\Sigma_-=O(t^{-\frac{1}{2}})\\
N_2-\frac{3}{4}=O(t^{-\frac{1}{2}})\\
N_3+\frac{3}{4}=O(t^{-\frac{1}{2}})
\end{eqnarray*}
With the same procedure as in the Bianchi II case, using now the Ellis-MacCallum solution we arrive at:

\begin{eqnarray*}
P=O(t^{-\frac12})
\end{eqnarray*}

\section{The bootstrap argument}
The argument which will lead us to our main conclusions is a bootstrap argument, a kind of continuous induction argument. The argument will
work as follows (see 10.3 of \cite{RA} for a detailed discussion). One has a solution of the evolution equations and assumes that the norm of
that function depends continuously on the time variable. Assuming that one has small data initially at $t_0$, i.e. the norm of our function
is small, one has to improve the decay rate of the norm such that the assumption that $[t_0,T)$ with $T<\infty$ is the maximal interval
 on which a solution with bounded norm corresponding to the prescribed initial data exists would lead to a contradiction. This is a way to obtain global existence for small data. In our case global existence is
already clear but if the argument works we also obtain information about how the solution behaves asymptotically which is our goal.
 The interval we look at is $[t_0,t_1)$ and we will present the estimates assumed in the following for the different cases. 
All prefactors on the right hand side are positive and as small as we want.

\subsection{Bootstrap assumptions}
A first task is to find the suitable bootstrap assumptions. We choose a slightly slower decay for the anisotropy and the curvature
 variables than in the linearized cases with the hope that using the central equations, we are able to obtain the same decay as in the linearized
 case. For the estimate of $P$ we start with a slower decay than the ones obtained in section \ref{peta} as well. The assumption of small data
 here is in the sense that our solutions are not ``far away'' from our special solutions. In general to improve an estimate the corresponding
 evolution equation will be integrated. The assumptions made for the different Bianchi cases exclude the vacuum case, since the values of $\Omega$
 due to the constraint equation are near the corresponding values of $\Omega$ of the special solutions, thus far from being zero.
\subsubsection{Bootstrap assumptions for Bianchi II}
\begin{eqnarray*}
|\Sigma_+-\frac{1}{8}| &\leq& A_+ (1+t)^{-\frac{3}{8}}\\
|\Sigma_-|&\leq& A_-(1+t)^{-\frac{3}{4}}\\
 |N_1-\frac{3}{4}| &\leq& A_c (1+t)^{-\frac{3}{8}}\\
P&\leq& A_m (1+t)^{-\frac{1}{3}}
\end{eqnarray*}
\subsubsection{Bootstrap assumptions for Bianchi VI$_0$}
\begin{eqnarray*}
|\Sigma_++\frac{1}{4}| &\leq& A_+ (1+t)^{-\frac{3}{8}}\\
|\Sigma_-|&\leq &A_-(1+t)^{-\frac{3}{8}}\\
 |N_2-\frac{3}{4}|& \leq &A_{c1} (1+t)^{-\frac{3}{8}}\\
|N_3+\frac{3}{4}| &\leq& A_{c2} (1+t)^{-\frac{3}{8}}\\
P&\leq &A_m (1+t)^{-\frac{1}{3}}
\end{eqnarray*}

\subsection{Estimate of the mean curvature}
The first variable we estimate is the trace of the second fundamental form or equivalently the Hubble variable. Let us rewrite (\ref{H-1}):
\begin{eqnarray}\label{HH}
 \partial_t(H^{-1})=\frac{3}{2}+D
\end{eqnarray}
with

\begin{eqnarray*}
D=\frac{1}{12}(\frac{R}{H^2}+\frac{3}{H^2}\sigma_{ab}\sigma^{ab})+\frac{4\pi S}{3H^2}
\end{eqnarray*}
Integrating (\ref{H-1}) and since $t_0=\frac{2}{3}H^{-1}(t_0)$ (this choice was made in section (\ref{tc})):

\begin{eqnarray*}
 H(t)=\frac{1}{\frac{3}{2}t+I}=\frac{2}{3}t^{-1}\frac{1}{1+\frac{2}{3}It^{-1}}
\end{eqnarray*}
with
\begin{eqnarray*}
 I=\int^{t}_{t_0}D(s)ds
\end{eqnarray*}
Now for Bianchi II:
\begin{eqnarray*}
D_{II}=\frac{3}{2}(\Sigma_+^2+\Sigma_-^2)+\frac{4\pi S}{3H^2}-\frac{N_1^2}{24} 
\end{eqnarray*}
and for Bianchi VI$_0$:

\begin{eqnarray*}
D_{VI_0}=\frac{3}{2}(\Sigma_+^2+\Sigma_-^2)+\frac{4\pi S}{3H^2}-\frac{1}{24}(N_2-N_3)^2
\end{eqnarray*}
It turns out that in all cases $D$ is small, in particular from the different bootstrap assumptions we obtain for for Bianchi II and VI$_0$ respectively:

\begin{eqnarray}\label{DD}
|D| \leq \epsilon_{2/3}(1+t)^{-\frac{3}{8}}
\end{eqnarray}
with
\begin{eqnarray*}
&&\epsilon_2=C(A_+ +A_-^2+A_c+A_m^2)\\
&&\epsilon_3=C(A_++A_-^2+A_{c1}+A_{c2}+A_m^2)
\end{eqnarray*}
We arrive at:

\begin{eqnarray*}
\frac{2}{3}t^{-1}I=O(\epsilon_{2/3}t^{-\frac{3}{8}})
\end{eqnarray*}
The results for the Hubble variable is
\begin{eqnarray}\label{H}
\boxed{H=\frac{2}{3}t^{-1}(1+O(\epsilon_{2/3} t^{-\frac{3}{8}}))}
\end{eqnarray}

We also obtain an estimate for the determinant using the estimate of $H$
and integrating (\ref{det}) in both directions.

\begin{eqnarray}\label{esdet}
C(t_0) t^{4-\epsilon} \leq  g(t) \leq C(t_0) t^{4+\epsilon}
\end{eqnarray}
\subsection{Estimate of the metric}\label{MM}
Consider the following equation in the sense of \textit{components}:
\begin{eqnarray*}
\bar{g}^{ab}=t^{\frac{p}{q}}g^{ab}
\end{eqnarray*}
 In particular we will consider the components $g^{22}$ and $g^{33}$, which means that for Bianchi II $\frac{p}{q}=\frac32$ and for Bianchi VI$_0$ 
$\frac{p}{q}=1$. We will show that:

\begin{eqnarray*}
\frac{d}{dt}(t^{-\gamma}\bar{g}^{ab})=t^{-\gamma-1}\bar{g}^{ab}(-\gamma+\frac{p}{q}+\frac{\dot{g}^{ab}}{g^{ab}}t)\leq -\eta t^{-\gamma-1}\bar{g}^{ab}
\end{eqnarray*}
with $\eta$ positive with the help of the bootstrap assumptions and choosing $\gamma$ in a suitable way. This means then that:

\begin{eqnarray*}
\frac{d}{dt}(t^{-\gamma}\bar{g}^{ab})\leq 0
\end{eqnarray*}
which implies what we wanted to show:

\begin{eqnarray}\label{eta}
g^{ab}(t)\leq t_0^{-\gamma+\frac{p}{q}}g^{ab}(t_0)t^{-\frac{p}{q}+\gamma}.
\end{eqnarray}
For the covariant components one can do the same by defining $\bar{g}_{ab}=t^{-\frac{p}{q}}g_{ab}$. One obtains:

\begin{eqnarray*}
\frac{d}{dt}(t^{\gamma}\bar{g}_{ab})=t^{\gamma-1}\bar{g}_{ab}(\gamma-\frac{p}{q}+\frac{\dot{g}_{ab}}{g_{ab}}t)\geq \eta t^{\gamma-1}\bar{g}^{ab}
\end{eqnarray*}

For the last step one can actually use the same $\gamma$ as for the contravariant components since $\dot{g}_{ab}g^{ab}=-g_{ab}\dot{g}^{ab}$. 
In other words once (\ref{eta}) is shown, we also have:

\begin{eqnarray*}
g_{ab}(t)\leq t_0^{\gamma-\frac{p}{q}}g_{ab}(t_0)t^{\frac{p}{q}-\gamma}
\end{eqnarray*}
From the definitions made one can obtain:

\begin{eqnarray}\label{KK}
&&\dot{g}^{11}=2g^{11}H(-1+2\Sigma_+)\\\label{KK2}
&&\dot{g}^{22}=2g^{22}H(-1-\Sigma_+-\sqrt{3}\Sigma_-)\\\label{KK3}
&&\dot{g}^{33}=2g^{33}H(-1-\Sigma_++\sqrt{3}\Sigma_-)
\end{eqnarray}

Then we have with (\ref{H}) for the components $g^{22}$ and $g^{33}$:

\begin{eqnarray*}
\eta&=&\gamma+2Ht(1+\Sigma_+\pm\sqrt{3}\Sigma_-)-\frac{p}{q}\\
&=&\gamma+\frac{4}{3}(1+O(\epsilon_{2/3}t^{-\frac{3}{8}}))(1+\Sigma_+\pm\sqrt{3}\Sigma_-)-\frac{p}{q}
\end{eqnarray*}
In both Bianchi II and VI$_0$:

\begin{eqnarray*}
\frac{4}{3}(1+\Sigma_+)-\frac{p}{q}=O(A_+(1+t)^{-\frac{3}{8}})
\end{eqnarray*}
which enables us to choose $\gamma$ in such a way that $\eta$ is positive. 
Different values of $\Sigma_+$ correspond to different exponents in the components of the metric.
Using the estimates of $g^{22}$ and $g^{33}$ we obtain then the estimate for the other component
of the metric $g_{11}$ via the estimate of the determinant. We could also proceed directly from (\ref{KK}).

Summarizing this means that asymptotically up to a positive constant which depends only on $t_0$
the components (and their inverses) of the metrics $g_{II}$ for Bianchi II
 and $g_{VI_0}$ for Bianchi VI$_0$ have the same decay up to an $\epsilon$ as the
 corresponding components of the Collins-Stewart and Ellis-MacCallum solution respectively:

\begin{eqnarray*}
&&C(t_0)t^{-\epsilon}\leq \frac{g_{II}}{g_{CS}}\leq C(t_0)t^{+\epsilon}\\
&&C(t_0)t^{-\epsilon}\leq \frac{g_{VI_0}}{g_{EM}}\leq C(t_0)t^{+\epsilon}
\end{eqnarray*}

\subsection{Estimate of $P$}
We can express the derivative of the metric as follows:

\begin{eqnarray*}
\dot{g}^{bf}=2H(\Sigma^b_a-\delta^b_a)g^{af}
\end{eqnarray*} 

It follows from (\ref{charak}) and using (\ref{ha}):
\begin{eqnarray*}
\dot{V}=\dot{g}^{bf}V_bV_f=2H(\Sigma^b_a-\delta^b_a)g^{af}V_bV_f=2H(\Sigma^1_1g^{11}V_1^2+\Sigma^2_2g^{22}V_2^2+\Sigma^3_3g^{33}V_3^2)-2HV
\end{eqnarray*}
The maximum of $\Sigma^1_1$, $\Sigma^2_2$ and $\Sigma^3_3$ is for Bianchi II and VI$_0$ equal to $\frac{1}{4}+O(t^{-\frac38})$.
Thus:
\begin{eqnarray*}
\dot{V}\leq 2HV (-\frac34+\epsilon t^{-\frac38})
\end{eqnarray*}
Using now the estimate of $H$ and integrating :
\begin{eqnarray*}
V \leq V(t_0)(t/t_0)^{-1+\epsilon}
\end{eqnarray*}

from which follows:

\begin{eqnarray*}
P \leq P(t_0)(t/t_0)^{-\frac12+\epsilon}
\end{eqnarray*}

Choosing $P(t_0)\leq A_m t_0^{\frac12 -\epsilon}$ we arrive at:

\begin{eqnarray*}
P \leq A_m t^{-\frac12+\epsilon}
\end{eqnarray*}

which is an improvement of the bootstrap assumption and which has the consequence that:

\begin{eqnarray}\label{SH2}
 \boxed{\frac{S}{H^2} \leq C t^{-1+\epsilon}}
\end{eqnarray}
\subsection{Closing Bianchi II}
Until now we have estimates for $H$ and for $P$ in the interval $[t_0,t_1)$. We need to improve the other variables.
 Although Bianchi II and Bianchi VI$_0$ are more complicated, the main argument will be the same as in Bianchi I \cite{EN}. 
\subsubsection{Estimate for $\Sigma_-$}
\paragraph{Case I}
If $|\Sigma_-| \leq A_-(1+t)^{-1+\epsilon}$ holds there
 nothing more to do, since this is a better estimate then the one assumed.
\paragraph{Case IIa}
Assume now $\Sigma_- > A_-(1+t)^{-1+\epsilon}$.
 Define $t_2$ as the smallest number not smaller than $t_0$ with the property $\Sigma_- \geq A_-(1+t)^{-1+\epsilon}$.
Since we are assuming that $\Sigma_->0$ we can divide (\ref{3}) by $\Sigma_-$:
\begin{eqnarray*}
\frac{\dot{\Sigma}_-}{\Sigma_-}=&H[-(3+\frac{\dot{H}}{H^2})+\Sigma_-^{-1}\frac{8\pi}{2\sqrt{3}H^2}(S^2_2-S^3_3)]
\end{eqnarray*}
With (\ref{DD}),(\ref{H}), the fact that $S^2_2-S^3_3\leq S$, (\ref{SH2}) and our assumption:
\begin{eqnarray}\label{po}
 \frac{\dot{\Sigma}_-}{\Sigma_-}\leq -t^{-1}(1-\xi)
\end{eqnarray}
where $\xi$ is as small as we want. This variable $\xi$ contains a term of type $A_-^{-1}A_m^2$, but $A_-$ and $A_m$ can be chosen independently as small as needed. Also the $\epsilon$ coming from $S$ has to be chosen bigger than the $\epsilon$ coming from $\Sigma_-$. Note that it becomes clear here why we had to improve $P$ to arrive at (\ref{SH2}).
Integrating (\ref{po}) between $t_1$ and $t_2$:
\begin{eqnarray*}
 \Sigma_-(t_1)\leq \Sigma_-(t_2)t_2^{1-\xi}t_1^{-1+\xi}
\end{eqnarray*}
Assume $t_2=t_0$ then:
\begin{eqnarray*}
 \Sigma_-(t_1)\leq \Sigma_-(t_0)t_0^{1-\xi}t_1^{-1+\xi}\leq A_-t_1^{-1+\xi}
\end{eqnarray*}
since $\Sigma_-(t_0)$ can be chosen in such a way that the last inequality holds.
\paragraph{Case IIb}
If $t_2 >t_0$ then by continuity
 $\Sigma_-(t_2)\leq A_-(1+t)^{-1+\epsilon}$ which means that:
\begin{eqnarray*}
\Sigma_-(t_1)\leq A_- (1+t)^{\epsilon-\xi}t_1^{-1+\xi}\leq A_- t_1^{-1+\xi}
\end{eqnarray*}
if $\epsilon$ is chosen to be smaller than $\xi$. 
The argument for the case that $\Sigma_-$ is negative is the same, 
just define $\bar{\Sigma}_-=-\Sigma_-$ and use $S^3_3-S^2_2 \leq S$. This means that we could improve our bootstrap assumption to:
\begin{eqnarray*}
|\Sigma_-(t_1)| \leq A_- t_1^{-1+\xi}
\end{eqnarray*}

\subsubsection{Bootstrap assumptions for the other time variable}

We have found that for the estimates in the following section it was useful, although not essential, to use the other
 time variable $\tau$. Using the estimate of the Hubble variable (\ref{H}) in the definition of $\tau$ (\ref{tau}) we have:

\begin{eqnarray*}
 \tau-\tau_0=\int^t_{t_0} \frac{2}{3}t^{-1}(1+O(\epsilon t^{-\frac{3}{8}}))dt
\end{eqnarray*}
After integrating and observing that $\tau_0$ and $t_0$ are constants we arrive at:

\begin{eqnarray*}
t^{-\frac{2}{3}}=t_0^{-\frac23}e^{-\tau+\tau_0+\xi}
\end{eqnarray*}
where $\xi$ is small: $\xi=O(\epsilon (t^{-\frac38}+t_0^{-\frac38}))$.
So the bootstrap assumptions can be translated to the time variable $\tau$. We obtain:
\begin{eqnarray*}
|\tilde{\Sigma}_+ |&<&C A_+ e^{-\frac{9}{16} \tau}\\
|\tilde{\Sigma}_-| &<&C A_- e^{-\frac{9}{8}\tau}\\
|\tilde{N}_1| &<&C A_c e^{-\frac{9}{16}\tau}\\
 P &<&C A_m e^{-\frac{1}{2}\tau}
\end{eqnarray*}
Since we have an estimate of $H$ in both directions one can go also back from an estimate in terms of $\tau$ to an estimate of $t$ just
 by a multiplication by a constant which will not be relevant.

\subsubsection{Estimate for $\Sigma_+$ and $N_1$}

Define
\begin{eqnarray*}
\left(
\begin{matrix} \hat{\Sigma}_+ \\
\hat{N_1}
\end{matrix} \right)= \mathbf{M_{II}^{-1}}
\left(
\begin{matrix} \tilde{\Sigma}_+ \\
\tilde{N_1}
\end{matrix} \right)
\end{eqnarray*}
where $\mathbf{M_{II}}$ is the matrix of eigenvectors of the linearized system (\ref{II}). Then we have:
\begin{eqnarray*}
\left(
\begin{matrix} \hat{\Sigma}_+ \\
\hat{N_1}
\end{matrix} \right)'=-\frac{3}{4}\left(
\begin{matrix} 1&\sqrt{\frac{3}{2}} \\
-\sqrt{\frac{3}{2}}&1
\end{matrix} \right)\left(
\begin{matrix} \hat{\Sigma}_+ \\
\hat{N_1}
\end{matrix} \right)+O(A_m^2 e^{-\tau})\left(
\begin{matrix} 1\\
1
\end{matrix} \right)
\end{eqnarray*}
since $O(\tilde{\Sigma}_+^2+ \tilde{N}_1^2+ \tilde{\Sigma}_-^2+P^2)=O(A_m^2 e^{-\tau})$.
Multiplying the first equation by $\hat{\Sigma}_+$ and the second by $\hat{N}_1$ and adding both we obtain:

\begin{eqnarray*}
\frac{d}{dt}(\hat{\Sigma}_+^2+ \hat{N}_1^2)= -\frac{3}{2} (\hat{\Sigma}_+^2+\hat{N}_1^2)+(\tilde{\Sigma}_++\tilde{N}_1)O(A_m^2 e^{-\tau})
\end{eqnarray*}
\begin{eqnarray*}
\frac{d}{dt}[\log(\hat{\Sigma}_+^2+ \hat{N}_1^2)]= -\frac{3}{2}+(\tilde{\Sigma}_++\tilde{N}_1)(\hat{\Sigma}_+^2+ \hat{N}_1^2)^{-1}O(A_m^2 e^{-\tau})
\end{eqnarray*}
Let us assume now that:
\begin{eqnarray*}
 \tilde{\Sigma}_+^2 + \tilde{N}_1^2 > (A_+^2+A_c^2)e^{(-\frac{3}{2}+\xi)\tau}
\end{eqnarray*}
This implies:
\begin{eqnarray*}
 \hat{\Sigma}_+^2 + \hat{N}_1^2 > C(A_+^2+A_c^2)e^{(-\frac{3}{2}+\xi)\tau}
\end{eqnarray*}
\begin{eqnarray*}
\frac{d}{d\tau}[\log(\hat{\Sigma}_+^2+ \hat{N}_1^2)] \leq -\frac{3}{2}+\epsilon e^{(-\frac{1}{16}-\xi)\tau}
\end{eqnarray*}
From which follows that:
\begin{eqnarray*}
\hat{\Sigma}_+^2+ \hat{N}_1^2 &\leq& (\hat{\Sigma}_+^2(\tau_0)+ \hat{N}_1^2(\tau_0))e^{(-\frac{3}{2}+\epsilon)(\tau-\tau_0)}\\
                                &\leq& C(\hat{\Sigma}_+^2(t_0)+ \hat{N}_1^2(t_0))(\frac{t}{t_0})^{-1+\epsilon}
\end{eqnarray*}
{or:}
\begin{eqnarray*}
\tilde{\Sigma}_+^2+ \tilde{N}_1^2 \leq C(\tilde{\Sigma}_+^2(t_0)+ \tilde{N}_1^2(t_0))(\frac{t}{t_0})^{-1+\epsilon}
\end{eqnarray*}
Making now the same argument as in the end of the estimate of $\Sigma_-$ we arrive at improved estimates for $\tilde{\Sigma}_+$
 and $\tilde{N}_1$:
\begin{eqnarray*}
|\tilde{\Sigma}_+| \leq \tilde{\Sigma}_+(t_0)t^{-\frac{1}{2}+\epsilon}\\
|\tilde{N}_1| \leq \tilde{N}_1(t_0)t^{-\frac{1}{2}+\epsilon}
\end{eqnarray*}
i.e. for $\Sigma_+$ and $N_1$:
\begin{eqnarray*}
 |\Sigma_+-\frac{1}{8}|\leq A_+ (1+t)^{-\frac{1}{2}+\epsilon}\\
|N_1-\frac{3}{4}|\leq A_c (1+t)^{-\frac{1}{2}+\epsilon}
\end{eqnarray*}
We have closed now the bootstrap argument. Note that for this last improvement of the estimates $\Sigma_+$ and $N_1$
 we did not use the improved estimates for $\Sigma_-$ and $P$.

\subsection{Closing Bianchi VI$_0$}
This case is analogous to Bianchi II. The bootstrap assumptions with the variable $\tau$ read:
\begin{eqnarray*}
|\tilde{\Sigma}_+ |&<&C A_+ e^{-\frac{9}{16} \tau}\\
|\tilde{\Sigma}_-| &<&C A_- e^{-\frac{9}{16}\tau}\\
|\tilde{N}_2| &<&C A_{c1} e^{-\frac{9}{16}\tau}\\
|\tilde{N}_3| &<&C A_{c2} e^{-\frac{9}{16}\tau}\\
 P &<&C A_m e^{-\frac{1}{2}\tau}
\end{eqnarray*}
In this case what remains are the estimates for $\Sigma_+$, $\Sigma_-$, $N_2$ and $N_3$. In terms of the transformed linearization
\begin{eqnarray*}
\left(
\begin{matrix} \hat{\Sigma}_+ \\
\hat{\Sigma}_-\\
\hat{N_2}\\
\hat{N_3}
\end{matrix} \right)= \mathbf{M_{VI0}^{-1}}
\left(
\begin{matrix} \tilde{\Sigma}_+ \\
\tilde{\Sigma}_- \\
\tilde{N_2}\\
\tilde{N_3}
\end{matrix} \right)
\end{eqnarray*}
where $\mathbf{M_{VI0}}$ is the matrix of eigenvectors of the linearized system (\ref{VI0}) we have:
\begin{eqnarray*}
\left(
\begin{matrix} \hat{\Sigma}_+ \\
\hat{\Sigma}_-\\
\hat{N}_2 \\
\hat{N}_3
\end{matrix} \right)'=-\frac{3}{4}\left(
\begin{matrix} 1&-\sqrt{3}&0&0 \\
\sqrt{3}&1&0&0\\
0&0&1&-1\\
0&0&1&1
\end{matrix} \right)\left(
\begin{matrix} \hat{\Sigma}_+ \\
\hat{\Sigma}_-\\
\hat{N}_2 \\
\hat{N}_3
\end{matrix} \right)+O(A_m^2 e^{-\tau})\left(
\begin{matrix} 1\\
1\\
1\\
1
\end{matrix} \right)
\end{eqnarray*}
since $O(\tilde{\Sigma}_+^2+ \tilde{N}_2^2+ \tilde{N}_3^2+ \tilde{\Sigma}_-^2+P^2)=O(A_m^2 e^{-\tau})$.
As in the Bianchi II case, we arrive with the same procedure at
\begin{eqnarray*}
\frac{d}{dt}[\log(\hat{\Sigma}_+^2+ \hat{\Sigma}_-^2)]= -\frac{3}{2}+(\tilde{\Sigma}_++\tilde{\Sigma}_-)(\hat{\Sigma}_+^2+ \hat{\Sigma}_-^2)^{-1}O(A_m^2 e^{-\tau})
\end{eqnarray*}
\begin{eqnarray*}
\frac{d}{dt}[\log(\hat{N}_2^2+ \hat{N}_3^2)]= -\frac{3}{2}+(\tilde{N}_2+\tilde{N}_3)(\hat{N}_2^2+ \hat{N}_3^2)^{-1}O(A_m^2 e^{-\tau})
\end{eqnarray*}
and this means that:
\begin{eqnarray*}
\frac{d}{d\tau}[\log(\hat{\Sigma}_+^2+ \hat{\Sigma}_-^2)] \leq -\frac{3}{2}+\epsilon e^{(-\frac{1}{16}-\xi)\tau}
\end{eqnarray*}
and a similar expression for $N_2$ and $N_3$ such that in the end we arrive at the estimates we wanted to obtain. In this case as well it was not
necessary to use the improved estimate of $P$.

\subsection{Results of the bootstrap argument}
Since we have improved all estimates we have closed the bootstrap argument. Let us summarize the results obtained in this chapter in the following propositions:
\begin{prop}
Consider any $C^{\infty}$ solution of the Einstein-Vlasov system with reflection and Bianchi II symmetry and with $C^{\infty}$
 initial data. Assume that $|{\Sigma}_+(t_0)-\frac18|$, $|\Sigma_-(t_0)|$, $|N_1(t_0)-\frac{3}{4}|$ and $P(t_0)$ are sufficiently small. Then at
 late times the following estimates hold:
\begin{eqnarray*}
 H(t)&=&\frac{2}{3}t^{-1}(1+O(t^{-\frac{1}{2}+\epsilon}))\\
\Sigma_+-\frac{1}{8}&=&O(t^{-\frac{1}{2}+\epsilon})\\
\Sigma_-&=&O(t^{-1+\epsilon})\\
N_1-\frac{3}{4}&=&O(t^{-\frac{1}{2}+\epsilon})\\
P(t)&=&O(t^{-\frac{1}{2}+\epsilon})
\end{eqnarray*}
\end{prop}
\begin{prop}
Consider any $C^{\infty}$ solution of the Einstein-Vlasov system with reflection Bianchi VI$_0$ symmetry and with $C^{\infty}$
 initial data. Assume that $|\Sigma_+(t_0)+\frac{1}{4}|$, $|\Sigma_-(t_0)|$, $|N_2(t_0)-\frac{3}{4}|$, $|N_3(t_0)+\frac{3}{4}|$ and $P(t_0)$
 are sufficiently small. Then at late times the following estimates hold:
\begin{eqnarray*}
H(t)&=&\frac{2}{3}t^{-1}(1+O(t^{-\frac{1}{2}+\epsilon}))\\
\Sigma_++\frac{1}{4}&=&O(t^{-\frac{1}{2}+\epsilon})\\
\Sigma_-&=&O(t^{-\frac{1}{2}+\epsilon})\\
N_2-\frac{3}{4}&=&O(t^{-\frac{1}{2}+\epsilon})\\
N_3+\frac{3}{4}&=&O(t^{-\frac{1}{2}+\epsilon})\\
P(t)&=&O(t^{-\frac{1}{2}+\epsilon})
\end{eqnarray*}
\end{prop}

In the next chapter we will improve the estimates such that we can get rid of the $\epsilon$. However the results stated here in this section represent
in fact the core of our results.

\section{Main results}
\subsection{Arzela Ascoli}
Until now we have obtained estimates which show that the decay rates of the different variables
are up to an $\epsilon$ the decay rates one obtains from the linearization. We want to use the Arzela-Ascoli theorem. 
We will show the boundedness of the relevant variables and their derivatives. 
The variables $\Sigma_-$, $\Sigma_+$, $N_1$, $N_2$ and $N_3$ corresponding to the different Bianchi cases are bounded uniformly due
 to the constraint equation. In particular:
\begin{eqnarray*}
\Sigma_+^2 + \Sigma_-^2\leq 1 \\
N_1^2 \leq 12 \\
(N_2 -N_3)^2 \leq 12
\end{eqnarray*}
Note that $N_3$ is negative. The Hubble variable $H$ and its derivative is bounded as one can see from the estimates and \ref{H-1}.
We have also obtained with the bootstrap argument that $P$, which is non-negative, decays which means that $S/H^2$ is bounded. 
From the estimates obtained it is clear that $g^{ab}$ and its derivative are bounded. 
Now having a look at the central equations we see that the derivatives of $\Sigma_-$, $\Sigma_+$, $N_1$, $N_2$ and $N_3$
 are also bounded uniformly.
If we can bound the derivative of $S$ also the second derivatives of $\Sigma_-$, $\Sigma_+$, $N_1$, $N_2$, $N_3$ and $H$ are bounded.
For this purpose it is convenient to express the components of the energy momentum tensor in terms of integrals of the covariant momenta:

\begin{eqnarray*}
S_{ab}g^{ab}&=&\int f(t,p) p_a p_b g^{ab}(1+g^{cd}p_{c}p_d)^{-\frac{1}{2}}g^{-\frac{1}{2}}dp_{1}dp_{2}dp_{3}\\
&=&\int f(t,p) V(1+V)^{-\frac{1}{2}}g^{-\frac{1}{2}}dp_{1}dp_{2}dp_{3}
\end{eqnarray*}
The only term of the time derivative of $S$  which could cause problems is the time derivative of the distribution function,
 since $\dot{V}$ and $\dot{g}^{ab}$ can be bounded by $V$ and $g^{ab}$ respectively and we know that $S$ itself is bounded since $S/H^2$ is.
 The term with the time derivative of the distribution function can be handled with the Vlasov equation:
\begin{eqnarray*}
\int \dot{f}(t,p) V(1+V)^{-\frac{1}{2}}g^{-\frac{1}{2}}dp_{1}dp_{2}dp_{3}\\
=-\int (p^0)^{-1}C^d_{ba}p^{b}p_{d}\frac{\partial f}{\partial p_a} V(1+V)^{-\frac{1}{2}}g^{-\frac{1}{2}}dp_{1}dp_{2}dp_{3}\\
\end{eqnarray*}
Integrating by parts we obtain a term which can be bounded by $S$. Note that the momenta grow in the worst case with $t^{\gamma}$
 and that $p_0$ is also bounded from below since the particles are assumed to have mass. Now all the relevant quantities are bounded. Let $\{t_n\}$
 be a sequence tending to infinity and let $(\Sigma_-)_n(t)=\Sigma_-(t+t_n)$, $(\Sigma_+)_n(t)=\Sigma_+(t+t_n)$,
 $(N_1)_n(t)=N_1(t+t_n)$, $(N_2)_n(t)=N_2(t+t_n)$, $(N_3)_n(t)=N_3(t+t_n)$, $H_n(t) = H(t+t_n)$ and $S_n(t)=S(t+t_n)$.
Using the bounds already listed, the Arzela-Ascoli theorem can be applied. This implies that, after passing to a subsequence,
 $(\Sigma_-)_n$, $(\Sigma_+)_n$ $(N_1)_n$, $(N_2)_n$, $(N_3)_n$, $H_{n}$ and $S_n$ converge uniformly on compact sets to a limit $(\Sigma_-)_{\infty}$, $(\Sigma_+)_{\infty}$ $(N_1)_{\infty}$, $(N_2)_{\infty}$, $(N_3)_{\infty}$, $H_{\infty}$ and $S_{\infty}$
 respectively. The first derivative of these variables converges to the corresponding derivative of the limits since we have been able to bound
 the derivative of $S$ in the last section. Going to this limit it is easy to see that the variable $D$ of equation (\ref{HH}) is zero and
 consequently:
\begin{eqnarray*}
 H_{\infty}=\frac{2}{3}t^{-1}
\end{eqnarray*}
From (\ref{KK})-(\ref{KK3}) we see that for Bianchi II and VI$_0$ we obtain the optimal decay rates for the metric and for its derivative. This implies that we obtain the optimal decay
 rates for $P$. Since $S/H^2$ is zero asymptotically we obtain the same estimates for $\Sigma_-$, $\Sigma_+$, $N_1$, $N_2$ and $N_3$
 as in the Einstein-dust case. Introducing this estimates in (\ref{HH}), we also obtain the optimal estimate for $H$. Let us summarize the estimates.
\subsection{Optimal estimates}
\begin{thm}
Consider any $C^{\infty}$ solution of the Einstein-Vlasov system with reflection and Bianchi II symmetry and with $C^{\infty}$
 initial data. Assume that $|{\Sigma}_+(t_0)-\frac18|$, $|\Sigma_-(t_0)|$, $|N_1(t_0)-\frac{3}{4}|$ and $P(t_0)$ are sufficiently small. Then at
 late times the following estimates hold:
\begin{eqnarray*}
 H(t)&=&\frac{2}{3}t^{-1}(1+O(t^{-\frac{1}{2}}))\\
\Sigma_+-\frac{1}{8}&=&O(t^{-\frac{1}{2}})\\
\Sigma_-&=&O(t^{-1})\\
N_1-\frac{3}{4}&=&O(t^{-\frac{1}{2}})\\
P(t)&=&O(t^{-\frac{1}{2}})
\end{eqnarray*}
\end{thm}
\begin{thm}
Consider any $C^{\infty}$ solution of the Einstein-Vlasov system with reflection Bianchi VI$_0$ symmetry and with $C^{\infty}$
 initial data. Assume that $|\Sigma_+(t_0)+\frac{1}{4}|$, $|\Sigma_-(t_0)|$, $|N_2(t_0)-\frac{3}{4}|$, $|N_3(t_0)+\frac{3}{4}|$ and $P(t_0)$
 are sufficiently small. Then at late times the following estimates hold:
\begin{eqnarray*}
H(t)&=&\frac{2}{3}t^{-1}(1+O(t^{-\frac{1}{2}}))\\
\Sigma_++\frac{1}{4}&=&O(t^{-\frac{1}{2}})\\
\Sigma_-&=&O(t^{-\frac{1}{2}})\\
N_2-\frac{3}{4}&=&O(t^{-\frac{1}{2}})\\
N_3+\frac{3}{4}&=&O(t^{-\frac{1}{2}})\\
P(t)&=&O(t^{-\frac{1}{2}})
\end{eqnarray*}
\end{thm}
For the cases Bianchi II and VI$_0$ we are also able to obtain the optimal estimate for the metrics:
\begin{cor}
Consider the same assumptions as in the previous theorem concerning Bianchi II and VI$_0$ respectively. Then
\begin{eqnarray*}
&&g_{II}=t\diag(K_1, t^{1/2}K_2, t^{1/2}K_3)\\
&&g_{VI_0}=t\diag (t K_4 ,K_5 ,K_6 )
\end{eqnarray*}
with $K_n=C_n+O(t^{-\frac{1}{2}})$ and where $C_1$-$C_6$ are independent of time. The corresponding
result for the inverse metric also holds.
\end{cor}
We see that the error in the metrics comes from the error in $\Sigma_+$.
\subsection{Kasner exponents}

From (\ref{TF}) we see that the eigenvalues (\ref{ev}) of the second fundamental form with respect to the induced metric are also
 the solutions of:
\begin{eqnarray*}
 \det(\sigma^i_j-[\lambda-\frac{1}{3}k]\delta^i_j)=0
\end{eqnarray*}
Let us define the eigenvalues of $\sigma_{ij}$ with respect to $g_{ij}$ by $\widehat{\lambda}_i$, we have that:
\begin{eqnarray*}
\widehat{\lambda}_i= \lambda_i-\frac{1}{3}k
\end{eqnarray*}
Note that $\Sigma_i (\widehat{\lambda}_i)^2=\sigma_{ab}\sigma^{ab}$. In the cases Bianchi II and VI$_0$ since everything is diagonal
 the Kasner exponents are easy to calculate. Using the optimal estimates for $\Sigma_+$, $\Sigma_-$ and $H$ and the fact that the sum
 of the generalized Kasner exponents is equal to one, we finally arrive at the generalized Kasner exponents for Bianchi II which are
 $(\frac{1}{4},\frac{3}{8},\frac{3}{8})$ and for Bianchi VI$_0$ $(\frac{1}{2},\frac{1}{4},\frac{1}{4})$ in both cases up to an error
 of order $O(t^{-\frac{1}{2}})$. Let us summarize these result:
\begin{cor}
Consider the same assumptions as in the previous theorem concerning Bianchi II and VI$_0$ respectively. Then:
\begin{eqnarray*}
 p_{II}= p_{CS}+O(t^{-\frac{1}{2}})\\
p_{VI_0}=p_{EM}+O(t^{-\frac{1}{2}})
\end{eqnarray*}
\end{cor}
We see that the error in the Kasner exponents comes from the error 
in $H$.

\subsection{Estimates of the energy momentum tensor}

Before coming to the estimates of the energy momentum tensor we show that for Bianchi II 
$V_2$ and $V_3$ become constants, something similar can be done for Bianchi VI$_0$. 
Define $E=g^{22} V_2^2 + g^{33} V_3^2$, then:
\begin{eqnarray*}
\dot{E}=\dot{g}^{22}V_2^2+\dot{g}^{33}V_3^2\leq 2H(-1-\Sigma_++\sqrt{3}|\Sigma_-|)E
\end{eqnarray*}
Integrating
\begin{eqnarray*}
 \log [E/E(t_0)] =-\frac{3}{2} \log t/t_0 + O(\epsilon (t^{-\frac{1}{2}}+t_0^{-\frac{1}{2}}))
\end{eqnarray*}
We have the following inequality for $E$:
\begin{eqnarray*}
 E \leq C t^{-\frac{3}{2}}
\end{eqnarray*}
Since the components of the metric $g^{22}$ and $g^{33}$ tend to the corresponding components
of the Collins-Stewart solution we see that $V_2$ and $V_3$ become constant asymptotically. 
The same is true in the case of the Ellis-MacCallum solution. Now since $f(t_0,p)$ has compact support
on $p$, we obtain that there exists a constant C such that:
\begin{eqnarray*}
 f(t,p)=0 \ \ \ |p_i| \geq C
\end{eqnarray*}
Let us denote by $\hat{p}$ the momenta in an orthonormal frame. Since $f(t,\hat{p})$ is constant
along the characteristics we have:
\begin{eqnarray*}
 |f(t,\hat{p})|\leq \Vert f_0 \Vert =\sup \{|f (t_0, \hat p)| \}
\end{eqnarray*}
Putting these facts together we arrive at the estimates which we summarize in the following:
\begin{cor}
Consider the same assumptions as in the previous theorem concerning Bianchi II. Then
\begin{eqnarray*}
&& \rho= \rho_{CS}(1+O(t^{-\frac12}))\\
&&S_{ij}\leq  C|f_0|t^{-3}
\end{eqnarray*}
\end{cor}
\begin{cor}
Consider the same assumptions as in the previous theorem concerning Bianchi VI$_0$ respectively. Then
\begin{eqnarray*}
 &&\rho=\rho_{EM}(1+O(t^{-\frac12}))\\
&&S_{ij}\leq  C|f_0|t^{-3}
\end{eqnarray*}
\end{cor}
The error in the energy density comes from the error in $H$.

 \emph{Remark}
From the corollaries one can estimate the quotient $S_{ij}/\rho$ which is $O(t^{-1})$.
That this quotient vanishes asymptotically means that the matter behaves as dust asymptotically as expected.

\section{Conclusions and Outlook}
The results concerning Bianchi II generalize the results obtained in \cite{RU}. For Bianchi VI$_0$ even for the reflection
 symmetric case there is no analogous previous result. The reason is that it is not compatible with the LRS-symmetry. 
Thus our result concerning Bianchi VI$_0$ shows clearly that the methods developed are powerful in the sense that one can obtain
 results which where out of reach with the techniques developed until now. However in contrast to \cite{RU} our results are restricted
 to the case of small data. In order to remove this assumption the Liapunov functions discovered for the fluid model
could be helpful. In the case of Bianchi II the future asymptotics are known globally even in the tilted case \cite{HL}.

We also hope to extend these results to the case without reflection symmetry in a future publication.

In our argument we have used the Arzela-Ascoli theorem, but only at the end. Thus there exist a lot of estimates
where one has control over the constants involved. Probably this could help for a numerical analysis of the Einstein-Vlasov equation which
is quite difficult.

Another path of generalizing our results could be the extension to higher dimensions. For the vacuum case geodesic completeness was shown
 for some homogeneous models in higher dimensions \cite{GOE}. The work on homogeneous spacetimes in higher dimensions may also shed some light
 on the inhomogeneous case in four spacetime dimensions.

\section*{Acknowledgments}
I would like to thank Alan Rendall for the continuous help and the transmission of a certain way of
 understanding mathematics. The point of view that mathematics with its rigour enables one to a have a deeper understanding of physical
 phenomena. This work has been funded by the Deutsche Forschungsgemeinschaft via the SFB 647-project B7.

\end{document}